\pdfoutput=1
\documentclass{SciPost}

\hypersetup{
    colorlinks,
    linkcolor={red!50!black},
    citecolor={blue!50!black},
    urlcolor={blue!80!black}
}
\usepackage[bitstream-charter]{mathdesign}
\DeclareSymbolFont{usualmathcal}{OMS}{cmsy}{m}{n}
\DeclareSymbolFontAlphabet{\mathcal}{usualmathcal}
\fancypagestyle{SPstyle}{
\fancyhf{}
\lhead{\colorbox{scipostblue}{\bf \color{white} ~SciPost Physics }}
\rhead{{\bf \color{scipostdeepblue} ~Submission }}

\fancyfoot[C]{\textbf{\thepage}}
}

\usepackage[T1]{fontenc}

\usepackage{booktabs}
\usepackage{xcolor}
\newcommand{\HH}{HH}
\newcommand{\bbbb}{b\bar{b}b\bar{b}}
\newcommand{\bbWW}{b\bar{b}W^+W^-}
\newcommand{\ttbar}{t\bar{t}}
\newcommand{\qscore}{q_{\mathrm{score}}}
\newcommand{\mlead}{m_{\mathrm{lead}}}
\newcommand{\msubl}{m_{\mathrm{subl}}}

\newcommand{\pt}{p_{\mathrm{T}}}
\newcommand{\HT}{H_{\mathrm{T}}}

\DeclareRobustCommand{\todo}[1][]{%
  \textcolor{red}{\textbf{[to-do}\if\relax\detokenize{#1}\relax\else\textbf{:} #1\fi\textbf{]}}}

\begin{document}

\pagestyle{SPstyle}

\begin{center}{\Large \textbf{\color{scipostdeepblue}{
Combinatorial Residual Network: a decay-chain-aware classifier without explicit jet assignment\\
}}}\end{center}

\begin{center}\textbf{
John Alison\textsuperscript{$\star$},
Patrick Bryant,
Alejandro Gomez Espinosa\textsuperscript{$\dagger$},
Christian Guinto-Brody,
Aniket Khanal,
Chuyuan Liu,
Sindhu Murthy and
Guanglei Zhao
}\end{center}

\begin{center}
Department of Physics, Carnegie Mellon University, Pittsburgh, PA, USA
\\[\baselineskip]
$\star$ \href{mailto:johnalison@cmu.edu}{\small johnalison@cmu.edu}\,,\quad
$\dagger$ \href{mailto:alejandrogomezespinosa@cmu.edu}{\small alejandrogomezespinosa@cmu.edu}
\\[\baselineskip]
{\small Authors are listed in alphabetical order.}
\end{center}

\section*{\color{scipostdeepblue}{Abstract}}
\textbf{\boldmath{%
Final states with several jets from the decays of heavy particles pose a combinatorial problem: the jets must be assigned, at least implicitly, to their parent particles.
The Combinatorial Residual Network (CoReNet) is a classifier that encodes the assumed decay chain in its architecture, building the event representation level by level along the decay hierarchy and combining all jet assignment hypotheses, evaluated with shared weights, with learned weights.
The exchange symmetries of the decay are built in, the network size does not depend on the number of hypotheses, and no generator-level jet assignments are used in training.
Developed for the $\bbbb$ final state and used in several CMS searches, CoReNet with 1508 parameters performs well above kinematic baselines on fast-simulated Higgs boson pair production against QCD multijet events and trains in minutes on a single GPU, and its hypothesis weights select the truth-matched jet pairing in most signal events in which it is defined.
Since only the encoded decay chain is specific to the final state, we test the construction on $\HH \to \bbWW$ against semileptonic $\ttbar$, where encoding the decay chain of the background as well improves the classification at a cost of about $10^{3}$ parameters.
In both final states, CoReNet performs comparably to a state-of-the-art transformer-based network more than two orders of magnitude larger.
The implementation is publicly available at \url{https://github.com/cms-cmu/CoReNet}.
}}

\vspace{\baselineskip}

\noindent\textcolor{white!90!black}{%
\fbox{\parbox{0.975\linewidth}{%
\textcolor{white!40!black}{\begin{tabular}{lr}%
  \begin{minipage}{0.6\textwidth}%
    {\small Copyright attribution to authors. \newline
    This work is a submission to SciPost Physics. \newline
    License information to appear upon publication. \newline
    Publication information to appear upon publication.}
  \end{minipage} & \begin{minipage}{0.4\textwidth}
    {\small Received Date \newline Accepted Date \newline Published Date}%
  \end{minipage}
\end{tabular}}
}}
}

\vspace{10pt}
\noindent\rule{\textwidth}{1pt}
\tableofcontents
\noindent\rule{\textwidth}{1pt}
\vspace{10pt}

\section{Introduction}
\label{sec:intro}

Final states with multiple jets from the decays of heavy particles appear throughout the physics program of the Large Hadron Collider (LHC), from top quark pair production and its association with Higgs bosons to multi-Higgs production and searches for new resonances decaying to quarks.
In all of them, the signal must be separated from a large QCD multijet background, and interpreting an event requires knowing, at least implicitly, which jets originate from which parent particle.
Since jets from different partons are difficult to distinguish beyond their $b$ tagging, this is a combinatorial problem whose size grows factorially with the number of jets.

Previous approaches to this problem have used machine learning to assign jets to their parent particles, using generator-level truth as the target, and then to classify signal against background~\cite{Shmakov:2021qdz,Fenton:2020woz,Fenton:2023jdb,Li:2024lml,Chiang:2025hhh,CMS:2026hhh}.
In a typical LHC analysis, however, the jet assignment is only an intermediate step.
The sensitivity is determined by how well the final discriminant separates signal from background.
Supervising the assignment is worthwhile only if learning the truth-matched pairing improves the classification, an assumption that is not usually tested directly.

In this paper we describe a different approach, in which the decay chain of the final state is encoded in the architecture of the network.
The Combinatorial Residual Network (CoReNet) builds its event representation following the decay hierarchy of the final state: jets are combined into candidates for the intermediate particles, and these into candidates for the full decay.
Jets are preselected by their $b$-tagging properties, every jet assignment hypothesis is evaluated, and the hypotheses are combined with learned weights using residual networks.
The same weights are applied to all objects at each level of the hierarchy, so the network has only $\mathcal{O}(10^{3})$ trainable parameters, and the exchange symmetries of the decay are built into the construction rather than learned.
CoReNet is trained with the classification loss alone and needs no generator-level jet assignment.

This architecture was first described publicly under the name Hierarchical Combinatorial Residual Network (HCR) in Ref.~\cite{Manole:2022bhs}, and has since been used in several CMS analyses~\cite{CMS:2024zzb,CMS:HIG24010}.
In Ref.~\cite{Manole:2022bhs} the network served as a density-ratio estimator for modeling the multijet background from data, learning the ratio between events with three and four $b$-tagged jets, and in the CMS analyses it served both in this role and as the signal-versus-background classifier.
As a density-ratio estimator it is trained without the four-tag events of the signal region and then applied to them, so its predictions are made out of sample.
Here we give a complete description of the architecture and study its classification performance.

We use Higgs boson pair production in the $\bbbb$ final state as the main benchmark, but the architecture can be applied to any final state with a known decay chain.
On fast-simulation samples against QCD multijet background, CoReNet performs well above kinematic baselines and comparably to a state-of-the-art reference, with more than two orders of magnitude fewer trainable parameters and a training time of minutes instead of hours.
Its performance is within 1\% of its best value already at a width of $d = 8$, i.e.\ with 1508 parameters, and its hypothesis weights select the truth-matched jet pairing in most signal events, although no assignment target is used in training.

To show that the construction is not specific to this final state, we adapt the hierarchy to $\HH \to \bbWW$ with one leptonically decaying $W$ boson, against a semileptonic $\ttbar$ background, where CoReNet can encode the decay chain of the main background as well as that of the signal.
There CoReNet again performs comparably to the reference, with more than two orders of magnitude fewer parameters.

The paper is organized as follows.
The CoReNet architecture and its adaptation to $\HH \to \bbWW$ are described in Sec.~\ref{sec:architecture}.
Section~\ref{sec:simulation} describes the simulated samples and the event selection, and Sec.~\ref{sec:setup} the training and evaluation of the classifiers.
The results for both final states are presented in Sec.~\ref{sec:results}.
Section~\ref{sec:conclusion} discusses the results and the use of CoReNet in CMS analyses, and summarizes the conclusions.

\section{The Combinatorial Residual Network}
\label{sec:architecture}

\begin{figure}[t]
  \centering
  \includegraphics[width=\textwidth]{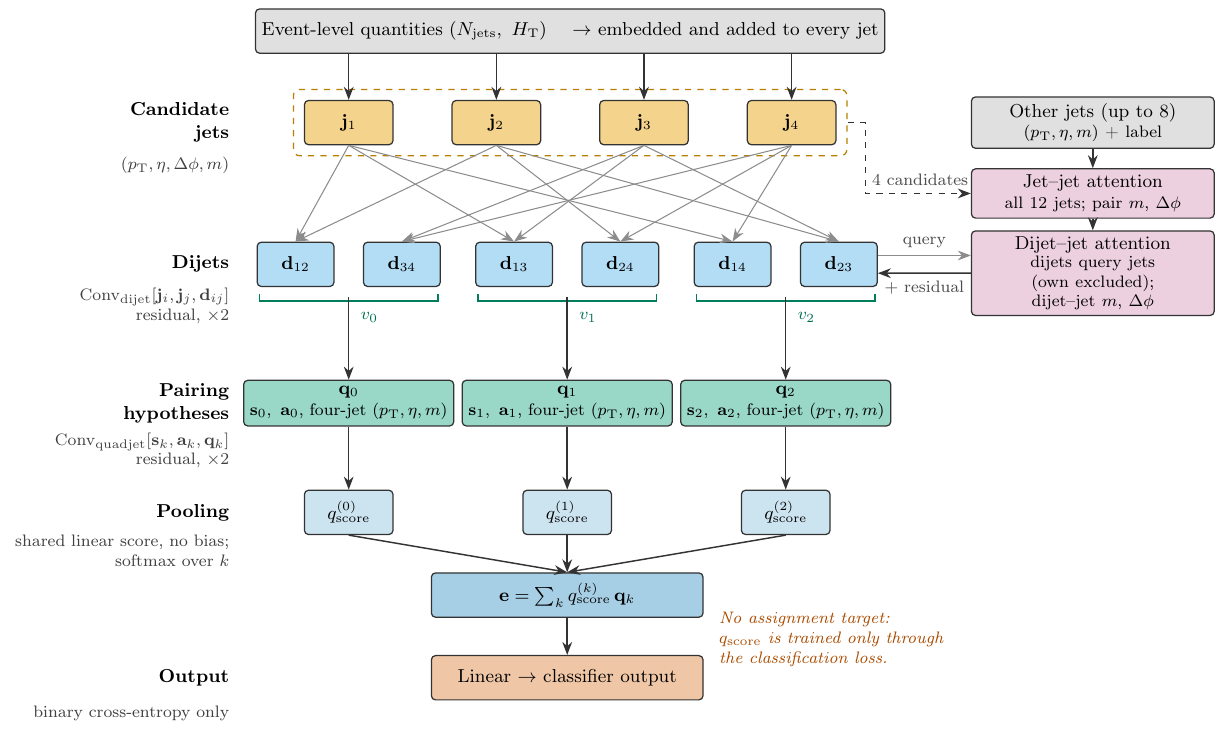}
  \caption{The CoReNet architecture for $\HH \to \bbbb$.
  The four candidate jets are combined into six dijets, which are grouped into the three pairing hypotheses $v_0$, $v_1$ and $v_2$.
  At each level, the representation of an object is built from the objects below it and from its own four-vector, and is refined by two residual layers with weights shared across objects.
  The additional jets enter through attention blocks that update the dijet representations.
  The hypotheses are pooled with the weights $\qscore$ into an event representation $\mathbf{e}$, which a linear layer maps to the classifier output.
  No assignment target is applied to $\qscore$.}
  \label{fig:architecture}
\end{figure}

CoReNet is built around the decay chain of the final state.
Each level of the chain is represented by a set of candidate objects formed from the previous level.
All possible assignments of objects to parent particles are enumerated, and the network evaluates every assignment hypothesis rather than choosing one.
The same weights are applied to every object at a given level, so the size of the network is set by the width of its internal representation and not by the number of objects or hypotheses.
The hypotheses are combined with learned weights, and the network is trained only on the classification objective without any truth-level assignment at any stage.

For example, in the $\HH \to \bbbb$ case, the chain has three levels: four jets form six dijets, and the dijets form three pairing hypotheses, each corresponding to one way of assigning the four jets to the two Higgs bosons (Fig.~\ref{fig:architecture}).
The network is described for this case in the following paragraphs as an illustration of the general design principle.
Other final states change the levels of the hierarchy but not the overall principle.

The inputs to the network are the four-momentum quantities $(\pt, \eta, \phi, m)$ of the final state objects.
In the case of $\HH \to \bbbb$, the $b$-tagging information is used only to choose the four candidate jets.
Additional jets, not chosen as candidates, are included as separate inputs to the network.
Finally, event-level quantities are provided as ancillary features: in this study, the number of selected jets and $\HT$.

Two transformations remove symmetries that the network would otherwise have to learn.
Each event is reflected about the transverse plane, $\eta \to -\eta$ for all objects, if needed so that the first candidate jet in the ordering of Sec.~\ref{sec:simulation} has $\eta > 0$.
Azimuthal angles enter only as absolute differences $|\Delta\phi|$ between objects, so the network is invariant under rotations about the beam axis and under $\phi \to -\phi$.
These and all other input transformations are applied inside the network, so no external standardization of the inputs is needed (App.~\ref{app:corenet}).
Unless stated otherwise, all layers are one-dimensional convolutions whose weights are shared across objects, and the nonlinearity is the SiLU function~\cite{Hendrycks:2016qxa,Elfwing:2017fzb}, denoted $\sigma$.

Each jet enters once for every dijet it belongs to, with its azimuth measured relative to that dijet, and each dijet is described by the kinematics of its four-vector sum (App.~\ref{app:corenet}).
These are mapped to $d$-dimensional embeddings by layers shared across all jets and all dijets, respectively.
The embedding of the two event-level quantities is added to every jet embedding.
We denote the resulting jet and dijet representations by $\mathbf{j}^{(0)}$ and $\mathbf{d}^{(0)}$.

The dijet representation is then refined by a convolution with kernel size and stride three, which reads the two jets and the dijet of each (jet, jet, dijet) slot:
\begin{equation}
  \mathbf{d}_{ij}^{(\ell+1)} = \sigma\!\left(\mathbf{d}_{ij}^{(0)} +
    \mathrm{Conv}_{\mathrm{dijet}}\big[\mathbf{j}_i^{(\ell)},\,
    \mathbf{j}_j^{(\ell)},\, \mathbf{d}_{ij}^{(\ell)}\big]\right),
  \qquad
  \mathbf{j}^{(\ell+1)} = \sigma\!\left(\mathbf{j}^{(0)} +
    \mathrm{Conv}_{\mathrm{jet}}\big[\mathbf{j}^{(\ell)}\big]\right).
  \label{eq:dijet-block}
\end{equation}

Two such layers are applied, $\ell = 0, 1$, and the jet representation is refined once between them, with its own skip connection to $\mathbf{j}^{(0)}$, so that the second layer reads the updated jets (App.~\ref{app:corenet}, Fig.~\ref{fig:dijet-block}a).

Each layer adds its output to the initial embedding of its level, as in residual networks~\cite{He:2015wf}.
The kinematic quantities built at each level therefore reach the next level directly, and each layer learns a correction to them rather than replacing them.
The two jets of each dijet enter in a canonical order, the jet that comes first in the candidate ordering of Sec.~\ref{sec:simulation} first, so the dijet representation is uniquely defined and the network can tell the two jets apart.
This stage is therefore deliberately not symmetric under exchange of the two jets.
The symmetry that matters for the pairing, exchange of the two dijets, is imposed exactly at the next level.

Jets beyond the four candidates are not part of any pairing hypothesis, but they carry information about initial- and final-state radiation and about decay products that were not selected as candidates.
They enter through two attention blocks~\cite{Vaswani:2017}: each jet first attends to all other jets, and each dijet then attends to the jets that are not its own, with the invariant mass and azimuthal separation of each pair modifying the attention (App.~\ref{app:corenet}).
The output is added to the dijet representation as a residual correction before the hypothesis level.
The attention blocks can be removed, so that the network uses only the candidate jets and the event-level quantities; for $\HH \to \bbbb$ at $d = 8$ this reduces the number of trainable parameters from 1508 to 782 (Sec.~\ref{sec:results}).

The six dijets are grouped into the three ways of splitting the four candidate jets into two pairs,
\begin{equation}
  v_0 = \{(12),(34)\},\quad
  v_1 = \{(13),(24)\},\quad
  v_2 = \{(14),(23)\}.
\end{equation}
Each is one hypothesis for which jets came from which Higgs boson.
We denote by $\tilde{\mathbf{d}}^{(0)}$ the dijet representation entering this level, which includes the correction from the attention blocks.
For a hypothesis $k$ with dijets $\tilde{\mathbf{d}}_a$ and $\tilde{\mathbf{d}}_b$, the two Higgs candidates are combined through their mean and their absolute difference,
\begin{equation}
  \mathbf{s}_k^{(\ell)} = \tfrac{1}{2}\big(\tilde{\mathbf{d}}_a^{(\ell)} + \tilde{\mathbf{d}}_b^{(\ell)}\big),
  \qquad
  \mathbf{a}_k^{(\ell)} = \tfrac{1}{2}\big|\tilde{\mathbf{d}}_a^{(\ell)} - \tilde{\mathbf{d}}_b^{(\ell)}\big| ,
  \label{eq:projections}
\end{equation}
which carry no trainable weights.
Each hypothesis also receives the four-vector of the four-jet system, which is the same for all three hypotheses, and its $(\log(1+\pt), \eta, \log(1+m))$ are embedded into $\mathbf{q}_k^{(0)}$.
The hypothesis representation is then refined in the same way as the dijets,
\begin{equation}
  \mathbf{q}_{k}^{(\ell+1)} = \sigma\!\left(\mathbf{q}_{k}^{(0)} +
    \mathrm{Conv}_{\mathrm{quadjet}}\big[\mathbf{s}_k^{(\ell)},\,
    \mathbf{a}_k^{(\ell)},\, \mathbf{q}_{k}^{(\ell)}\big]\right),
  \label{eq:quadjet-block}
\end{equation}
for two layers (Fig.~\ref{fig:dijet-block}b).
Between them, the dijets are refined once by a size-1 convolution with a skip connection to $\tilde{\mathbf{d}}^{(0)}$, as the jets are at the dijet level, so that the second layer reads $\tilde{\mathbf{d}}^{(1)}$.
Since $\mathbf{s}_k$, $\mathbf{a}_k$ and the four-jet kinematics are all unchanged under $\tilde{\mathbf{d}}_a \leftrightarrow \tilde{\mathbf{d}}_b$, the representation of each hypothesis is invariant under exchange of the two Higgs candidates.
The symmetry is built into the construction and costs no parameters.

CoReNet does not select a pairing.
A linear layer without bias, shared across hypotheses, assigns each hypothesis a score $g(\mathbf{q}_k)$, and the scores are normalized across the three hypotheses,
\begin{equation}
  \qscore = \mathrm{softmax}\!\left(
    \left[\,g(\mathbf{q}_0),\, g(\mathbf{q}_1),\, g(\mathbf{q}_2)\,\right]\right),
  \qquad
  \mathbf{e} = \sum_{k=0}^{2} \qscore^{(k)} \, \mathbf{q}_k .
  \label{eq:pooling}
\end{equation}
A final linear layer maps the event representation $\mathbf{e}$ to the classification output, and the network is trained with a classification loss alone, the binary cross-entropy for $\HH \to \bbbb$.

The weights $\qscore$ are not trained to identify a particular pairing.
Instead, the event representation $\mathbf{e}$ draws on all three hypotheses at once, with weights that the network sets event by event, so the classification can use features from every interpretation of the event simultaneously.
Each hypothesis, truth-matched or not, can carry information that separates signal from background, and this is the central idea of CoReNet: the classification does not depend on first choosing one interpretation of the event.
Choosing one would in any case rest on a convention, since at detector level a correct pairing is not well defined: it is obtained by matching generator-level quarks to reconstructed jets, and in background events it does not exist at all.
As a result, no truth-level information enters the training at any stage, and the network can be trained on any labeled sample, including samples taken from data, as in its use as a density-ratio estimator~\cite{Manole:2022bhs,CMS:2024zzb,CMS:HIG24010}.
Whether the learned weights nevertheless select the truth-matched pairing is studied in Sec.~\ref{sec:results}.

Because weights are shared across objects at every level, the number of trainable parameters depends only on the width $d$ of the internal representation.
With $d = 8$, the configuration used for the main results, CoReNet has 1508 trainable parameters.
For comparison, the SPANet configuration used as a reference (Sec.~\ref{sec:setup}) has 559846.

The construction described above for $\HH \to \bbbb$ follows from a small set of rules that apply to any final state with a known decay chain.
The decay chain is written as a tree whose leaves are the reconstructed objects and whose internal nodes are the intermediate particles.
Each internal node becomes one level of the network, with candidates whose four-vector is the sum of those of their decay products.
Objects of different types, such as jets, leptons or the missing transverse momentum, enter through their own embeddings, which also receive any features beyond the four-momentum, such as the lepton flavor.
The candidates of each node are enumerated from its decay products, with every reconstructed object used at most once per candidate, and the possible assignments of the full decay tree form the hypotheses, which are pooled with learned weights as in Eq.~\eqref{eq:pooling}.
When the two decay products of a node are identical composite objects, such as the two Higgs boson candidates in $\HH \to \bbbb$, they are combined through the mean and absolute difference of Eq.~\eqref{eq:projections}, so that their exchange symmetry is exact; otherwise their representations are read in a fixed order by the residual layer of the node.
The decay chains of several processes can be described on the same objects, sharing the candidates they have in common, so that the network evaluates background hypotheses alongside the signal hypothesis.
Every level is refined by residual blocks of the form of Eq.~\eqref{eq:dijet-block}, so the number of trainable parameters remains set by the width $d$.

\begin{figure}[t]
  \centering
  \includegraphics[width=\textwidth]{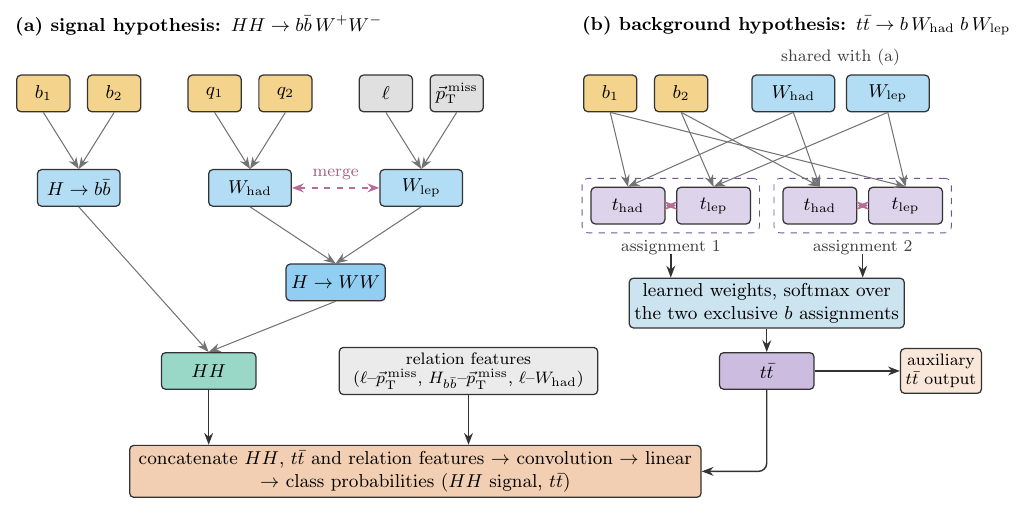}
  \caption{The CoReNet hierarchy for $\HH \to \bbWW$ with one leptonically decaying $W$ boson.
  (a) Signal hypothesis, used in both CoReNet$_{\HH}$ and CoReNet$_{\HH+\ttbar}$: the two leading $b$-tagged jets, $b_1$ and $b_2$, form the $H \to b\bar{b}$ candidate, the two leading other jets $q_1$ and $q_2$ the hadronic $W$ candidate, and the lepton and missing transverse momentum the leptonic $W$ candidate.
  (b) Background hypothesis, added in CoReNet$_{\HH+\ttbar}$: the two exclusive assignments of the $b$ jets to a hadronic and a leptonic top quark candidate are evaluated and combined with learned weights, as the three pairings are for $\bbbb$.
  Dashed arrows indicate the pairwise merge that connects the two halves of each hypothesis.
  In CoReNet$_{\HH+\ttbar}$ the two hypotheses and a set of relation features are combined into the classifier output, and the $\ttbar$ hypothesis also feeds an auxiliary output.}
  \label{fig:bbww-hierarchy}
\end{figure}

We apply these rules to Higgs boson pair production in the $\bbWW$ final state with one leptonically decaying $W$ boson, against a semileptonic $\ttbar$ background.
This final state differs from $\bbbb$ in the ways that matter most for the architecture: the two Higgs bosons decay differently, the final state contains a lepton and missing transverse momentum, and the dominant background has a decay chain of its own on the same objects.
The two leading $b$-tagged jets form the $H \to b\bar{b}$ candidate and the two leading remaining jets the hadronic $W$ candidate, so the signal hypothesis has a single assignment.
The leptonic $W$ candidate is built from the lepton and the missing transverse momentum, treated as a massless vector in the transverse plane, with no constraint on the $W$ mass.
The two $W$ candidates, merged with each other, form the $H \to WW$ candidate, and the two Higgs candidates, which are not identical and are therefore read in a fixed order, form the signal hypothesis (Fig.~\ref{fig:bbww-hierarchy}a).
In the configuration denoted CoReNet$_{\HH}$, the classifier output is computed from this hypothesis and from three pairwise relation features (App.~\ref{app:corenet}).

CoReNet$_{\HH+\ttbar}$ also encodes the decay chain of the $\ttbar$ background, reusing the two $W$ candidates: each $b$ jet is combined with one $W$ candidate into a hadronic and a leptonic top quark candidate, which are merged as the $W$ candidates are, and the two assignments of the $b$ jets to the top quarks, combined with learned weights as the three pairings are for $\bbbb$, form the background hypothesis (Fig.~\ref{fig:bbww-hierarchy}b).
Its classifier output is computed from the signal and background hypotheses and the relation features, and the background hypothesis also feeds an auxiliary output.
CoReNet$_{\HH}$ is trained with the binary cross-entropy, as for $\bbbb$, and CoReNet$_{\HH+\ttbar}$ with the cross-entropy over the signal and $\ttbar$ classes plus, with equal weight, the binary cross-entropy of the auxiliary output against the $\ttbar$ label; in both, the signal-class probability is the discriminant.
With $d = 8$, CoReNet$_{\HH}$ and CoReNet$_{\HH+\ttbar}$ have 2550 and 3832 trainable parameters.
Comparing the two configurations measures what encoding the decay chain of the background adds to the classification.

\section{Simulated samples and event selection}
\label{sec:simulation}

Signal and background samples are simulated for proton-proton collisions at $\sqrt{s} = 13$~TeV.
The signal is standard model $\HH$ production through gluon-gluon fusion, generated at next-to-leading order with POWHEG-BOX~\cite{Nason:2004rx,Frixione:2007vw,Alioli:2010xd,Heinrich:2017kxx,Heinrich:2019bkc} and showered with Pythia~8~\cite{Sjostrand:2014zea}, with both Higgs bosons decaying to $b\bar{b}$.
The background is QCD multijet production generated with Pythia~8 with $\hat{p}_{\mathrm{T}} > 75$~GeV.
It is filtered at generator level to contain at least four $b$ hadrons within $|\eta| < 2.5$, a generator-level $\HT > 350$~GeV, and jet $\pt$ requirements that approximate a multijet trigger.
This filter enriches the background in genuine four-$b$-quark configurations, which makes it considerably harder to separate from the signal than inclusive multijet production.
The samples contain 50000 generated signal events and 500000 background events after the generator-level filter.

For the $\bbWW$ final state, the signal is gluon-gluon fusion $\HH$ production generated at leading order with MadGraph5\_aMC@NLO 3.5.6~\cite{Alwall:2014hca} and showered with Pythia~8.
In Pythia the Higgs boson decays are restricted to $b\bar{b}$ and $WW$, and the events are filtered at generator level to one $H \to b\bar{b}$ and one $H \to WW \to \ell\nu q\bar{q}'$ decay, with $\ell = e$ or $\mu$, both $W$ charges, and a generator-level lepton $\pt > 20$~GeV.
The background is semileptonic $\ttbar$ production generated with Pythia~8 ($gg$ and $q\bar{q} \to \ttbar$) with the same decay-mode filter.
The samples contain 269591 signal and 3000000 background events after these generator-level filters; the signal sample is obtained from $3.2 \times 10^{6}$ generated $\HH$ events, of which 370735 pass the decay-mode filter and 269591 also the lepton $\pt$ requirement.
All of the generated samples used in this paper have been shared publicly~\cite{Khanal:2026data}.

The detector response is simulated with Delphes~3.5.1~\cite{deFavereau:2013fsa} using the CMS-like card of the JetClass-II dataset~\cite{Li:2024JetClassII}, without pileup.
Jets are clustered from PUPPI-weighted~\cite{Bertolini:2014bba} particle-flow candidates with the anti-$k_{\mathrm{T}}$ algorithm~\cite{Cacciari:2008gp}, as implemented in FastJet~\cite{Cacciari:2011ma}, with a distance parameter of $R = 0.4$.
The $b$ tagging is applied through the efficiency and misidentification parameterization of the card, which gives a binary tag for each jet.
Lepton efficiencies follow the parameterization of the card, which is based on CMS Run~2 measurements.

For the $\bbbb$ final state, jets are required to have $\pt > 20$~GeV and $|\eta| < 2.5$, and to be matched within $\Delta R < 0.4$ to a generator-level jet satisfying the same requirements.
Events are required to contain at least four selected jets, and up to twelve jets are kept per event; 36283 signal and 417978 background events pass this selection.
The jets are ordered with the $b$-tagged jets first, each group ordered in $\pt$, and the first four are taken as the candidate jets from which CoReNet builds its dijets and pairing hypotheses (Sec.~\ref{sec:architecture}).
The remaining jets, up to eight, are the additional jets that enter through the attention blocks.

For the $\bbWW$ final state, events are required to contain exactly one tight lepton: a muon with $\pt > 26$~GeV and $|\eta| < 2.4$, or an electron with $\pt > 35$~GeV and $|\eta| < 2.5$, with a loose isolation requirement.
Events with an opposite-sign same-flavor pair of loose leptons with $|m_{\ell\ell} - 91.2~\mathrm{GeV}| < 10$~GeV, or with a hadronic $\tau$ candidate, are rejected.
At least four jets with $\pt > 25$~GeV, $|\eta| < 2.4$ and $\Delta R > 0.4$ from the lepton are required, matched to generator-level jets as for $\bbbb$, at least two of which are $b$-tagged.
After this selection the samples contain 17511 signal and 274774 $\ttbar$ events.
The jets are ordered in the same way as for $\bbbb$, and the candidate objects of the two CoReNet configurations are built from them as described in Sec.~\ref{sec:architecture}.

\section{Training and evaluation}
\label{sec:setup}

The events are divided at random into five folds of equal size.
For each fold used as the test sample, three folds are used for training and one for validation, and the procedure is repeated so that every event is used for testing once.
The results quoted are the mean and sample standard deviation over the five test folds.
All classifiers are trained and evaluated on the same events with the same folds.
Event weights are normalized so that the signal and background weights each sum to one in the training sample, so both classes contribute equally to the loss, and the AUC is computed with the same weights.

CoReNet is trained for 20 epochs with the Adam optimizer~\cite{Kingma:2014}, following the recipe of its CMS deployments~\cite{CMS:2024zzb,CMS:HIG24010}, which is given in App.~\ref{app:corenet}.
The model after the best validation epoch is used.

To establish a baseline, two classifiers on eleven hand-engineered event-level observables are trained: a gradient-boosted decision tree (GBDT)~\cite{Friedman:2001} and a multilayer perceptron (MLP)~\cite{Rumelhart:1986}.
The observables are the masses $\mlead$ and $\msubl$ of the leading and subleading dijets, where the leading dijet is the one with the larger scalar sum of jet $\pt$, the four-jet mass $m_{4j}$, the angular separation $\Delta R$ between the two jets of each dijet, the number of selected jets, $\HT$, and the transverse momenta of the four candidate jets.
They are computed from the four candidate jets, paired into two dijets by the pairing whose dijet masses lie closest to the reconstructed Higgs boson mass peaks, measured as $\sqrt{[(\mlead - \mu_{1})/\sigma_{1}]^{2} + [(\msubl - \mu_{2})/\sigma_{2}]^{2}}$ with the peak positions $\mu_{1} = 109$~GeV and $\mu_{2} = 96$~GeV and widths $\sigma_{1} = 12$~GeV and $\sigma_{2} = 14$~GeV taken from Gaussian fits to the dijet mass distributions of truth-matched signal.
The GBDT is the histogram-based implementation of scikit-learn~\cite{Pedregosa:2011}, following LightGBM~\cite{Ke:2017}, and uses 300 boosting iterations with a learning rate of 0.05 and at most 31 leaves per tree.
The MLP has two hidden layers of 64 units each with ReLU activations, takes standardized inputs, and is trained with Adam~\cite{Kingma:2014} at a learning rate of $10^{-3}$ for up to 200 epochs with early stopping on a held-out tenth of the training events.

As a state-of-the-art reference we use SPANet~\cite{Shmakov:2021qdz,Fenton:2020woz,Fenton:2023jdb}, a transformer~\cite{Vaswani:2017} with symmetry-preserving attention that respects the invariances of the jet assignment problem.
Its architecture and training settings follow Ref.~\cite{Li:2024lml} and are described in App.~\ref{app:spanet}.
SPANet is trained jointly to assign jets to their parent particles and to classify signal against background, minimizing $\lambda_{\mathrm{assign}}\mathcal{L}_{\mathrm{assign}} + \lambda_{\mathrm{det}}\mathcal{L}_{\mathrm{det}} + \lambda_{\mathrm{cls}}\mathcal{L}_{\mathrm{cls}}$, where $\mathcal{L}_{\mathrm{assign}}$ is a jet-assignment loss with the truth-matched jets as targets, $\mathcal{L}_{\mathrm{det}}$ a detection loss on whether each Higgs boson can be reconstructed from the input jets, and $\mathcal{L}_{\mathrm{cls}}$ the signal-versus-background cross-entropy.
The truth-matched targets exist only for the $45.0\%$ of the selected signal events in which all four $b$ quarks are matched to distinct jets, and CoReNet does not use them at any stage.
Since the relative weights of the losses are a free choice, we train four configurations: $\lambda_{\mathrm{assign}} = 20$, 5 and 1 with $\lambda_{\mathrm{det}} = \lambda_{\mathrm{cls}} = 1$, the first being the weights of Ref.~\cite{Li:2024lml}, and the classification loss alone ($\lambda_{\mathrm{assign}} = \lambda_{\mathrm{det}} = 0$, $\lambda_{\mathrm{cls}} = 1$).
Unlike CoReNet, SPANet receives the $b$ tag of every jet as an input feature.

For $\bbWW$, the folds, event weights and training procedures of all classifiers are the same as for $\bbbb$.
The kinematic baselines are trained on 21 observables: the $\pt$, $\eta$ and $b$ tag of the four leading jets, the lepton $\pt$ and $\eta$, the missing transverse momentum, $\HT$, the invariant masses of the two $b$-tagged jets and of the two leading remaining jets, the transverse mass of the lepton and missing momentum, and the numbers of jets and of $b$-tagged jets.
For SPANet, the truth-matched targets exist for only $21.1\%$ of the selected signal events; their definition is given in App.~\ref{app:spanet}.

\section{Results}
\label{sec:results}

The results for both final states are presented together.
We first compare CoReNet with the kinematic baselines and with SPANet, and then study how its performance depends on the width of the network and on encoding the decay chain of the background.
Finally, we examine the hypothesis weights and the training time.

\begin{table}[t]
  \centering
  \caption{Five-fold classification performance (ROC AUC weighted by the
  class-normalized event weights). Uncertainties are standard deviations
  across folds. $p$-values are two-sided paired $t$-tests
  against CoReNet ($d=8$) across the five folds, with the Nadeau--Bengio
  variance correction for the overlap of the training sets~\cite{Nadeau:2003}. SPANet uses the architecture of
  Ref.~\cite{Li:2024lml} trained for 500 epochs, with $\lambda_{\mathrm{det}} = \lambda_{\mathrm{cls}} = 1$ and the
  assignment-loss weight $\lambda_{\mathrm{assign}}$ given in the table ($\lambda_{\mathrm{det}} = 0$ for the
  classification-only row); SPANet rows use, for each fold, the epoch checkpoint with the best validation AUC.}
  \label{tab:main}
  \footnotesize
  \begin{tabular}{lrrc}
    \toprule
    Model & Parameters & ROC AUC & corrected $p$ \\
    \midrule
    Kinematic GBDT              & ---       & $0.8294 \pm 0.0028$ & $6 \times 10^{-5}$ \\
    Kinematic MLP               & 4993      & $0.8295 \pm 0.0041$ & $3 \times 10^{-4}$ \\
    \midrule
    \textbf{CoReNet} ($d=8$)        & \textbf{1508}  & $\mathbf{0.8664 \pm 0.0031}$ & --- \\
    CoReNet ($d=8$), no additional-jet attention & 782 & $0.8578 \pm 0.0013$ & $0.012$ \\
    \midrule
    SPANet, classification only ($\lambda_{\mathrm{assign}} = 0$) & 559846 & $0.8536 \pm 0.0032$ & $0.015$ \\
    SPANet, $\lambda_{\mathrm{assign}} = 1$ (best)                 & 559846 & $0.8593 \pm 0.0047$ & $0.14$ \\
    SPANet, $\lambda_{\mathrm{assign}} = 5$   & 559846 & $0.8437 \pm 0.0052$ & $0.014$ \\
    SPANet, $\lambda_{\mathrm{assign}} = 20$  & 559846 & $0.8261 \pm 0.0011$ & $5 \times 10^{-5}$ \\
    \bottomrule
  \end{tabular}
\end{table}

\begin{figure}[t]
  \centering
  \includegraphics[width=0.78\textwidth]{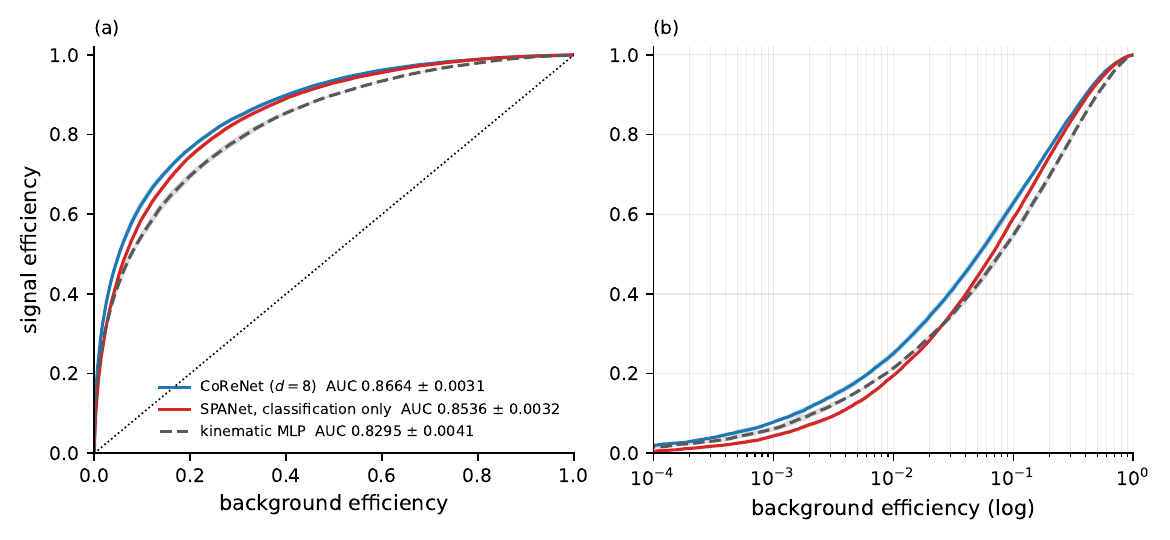}
  \caption{Five-fold mean ROC curves with cross-fold spread for CoReNet ($d=8$)
  (1508 parameters), the classification-only SPANet ($559846$ parameters, 500 epochs) and the kinematic MLP
  baseline, on (a) linear and (b) logarithmic background-efficiency axes.}
  \label{fig:roc}
\end{figure}

\begin{table}[t]
  \centering
  \caption{Five-fold classification performance on $\HH \to \bbWW$ against semileptonic $\ttbar$ (ROC AUC weighted by the class-normalized event weights, mean $\pm$ standard deviation over the five test folds).
  CoReNet$_{\HH}$ encodes the decay chain of the signal only, and CoReNet$_{\HH+\ttbar}$ also that of the $\ttbar$ background (Fig.~\ref{fig:bbww-hierarchy}); the auxiliary row uses the auxiliary $\ttbar$ output of CoReNet$_{\HH+\ttbar}$ as the discriminant, with score $1 - P(\ttbar)$.
  SPANet uses the architecture of Ref.~\cite{Li:2024lml} trained for 500 epochs, with $\lambda_{\mathrm{det}} = \lambda_{\mathrm{cls}} = 1$ and the assignment-loss weight given in the table ($\lambda_{\mathrm{det}} = 0$ for the classification-only row), and the checkpoint with the best validation AUC of each fold.
  The $p$-values are two-sided paired $t$-tests against CoReNet$_{\HH+\ttbar}$ across the five folds, with the Nadeau--Bengio correction~\cite{Nadeau:2003}.}
  \label{tab:bbww-hyp}
  \footnotesize
  \begin{tabular}{lrrc}
    \toprule
    Model & Parameters & ROC AUC & corrected $p$ \\
    \midrule
    Kinematic GBDT & --- & $0.8536 \pm 0.0041$ & $2 \times 10^{-5}$ \\
    Kinematic MLP  & 5633 & $0.8537 \pm 0.0019$ & $4 \times 10^{-7}$ \\
    \midrule
    CoReNet$_{\HH}$ & 2550 & $0.8859 \pm 0.0016$ & $3 \times 10^{-5}$ \\
    \textbf{CoReNet}$_{\HH+\ttbar}$ & \textbf{3832} & $\mathbf{0.9004 \pm 0.0018}$ & --- \\
    CoReNet$_{\HH+\ttbar}$, auxiliary $\ttbar$ output & 3832 & $0.8753 \pm 0.0005$ & $4 \times 10^{-5}$ \\
    \midrule
    SPANet, classification only ($\lambda_{\mathrm{assign}} = 0$) & 594520 & $0.9126 \pm 0.0025$ & $0.002$ \\
    SPANet, $\lambda_{\mathrm{assign}} = 1$  & 594520 & $0.8999 \pm 0.0039$ & $0.85$ \\
    SPANet, $\lambda_{\mathrm{assign}} = 5$  & 594520 & $0.8919 \pm 0.0034$ & $0.015$ \\
    SPANet, $\lambda_{\mathrm{assign}} = 20$ & 594520 & $0.8770 \pm 0.0078$ & $0.019$ \\
    \bottomrule
  \end{tabular}
\end{table}

\begin{figure}[t]
  \centering
  \includegraphics[width=0.95\textwidth]{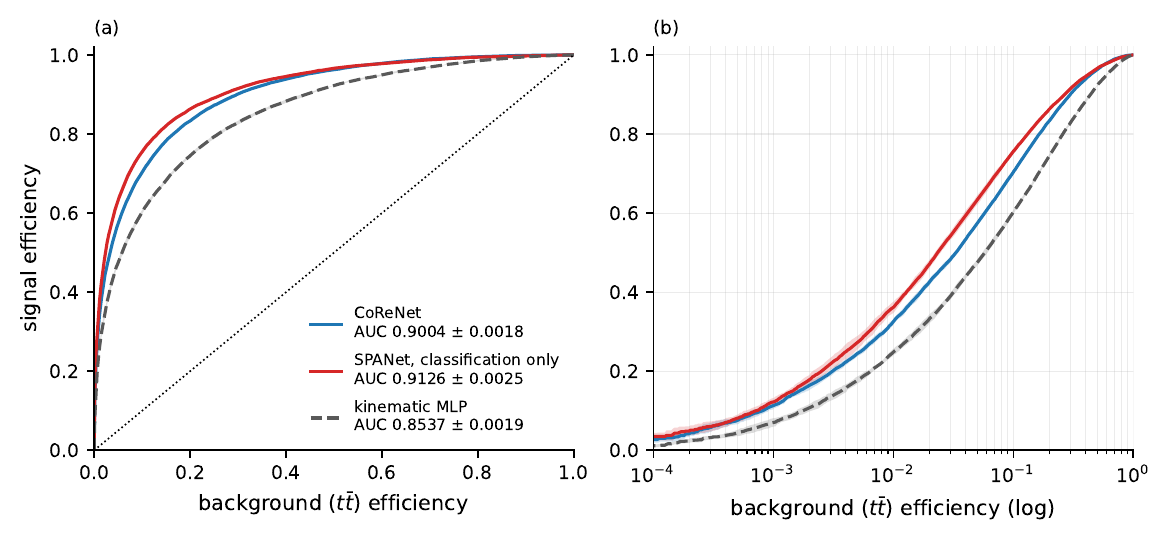}
  \caption{Five-fold mean ROC curves for $\HH \to \bbWW$ with cross-fold spread for CoReNet$_{\HH+\ttbar}$ (3832 parameters), SPANet trained with the classification loss alone (594520 parameters, 500 epochs) and the kinematic MLP baseline, on (a) linear and (b) logarithmic background-efficiency axes.
  SPANet uses, per fold, the checkpoint with the best validation AUC.}
  \label{fig:bbww-roc}
\end{figure}

Encoding the decay chain in the architecture captures more than a kinematic summary of the event.
The kinematic baselines are trained on observables that already encode the dijet masses and the angular structure of the signal (Sec.~\ref{sec:setup}), so a network can outperform them only by learning correlations beyond this summary.
In $\bbbb$, CoReNet reaches an AUC of $0.8664 \pm 0.0031$, against $0.8294 \pm 0.0028$ and $0.8295 \pm 0.0041$ for the GBDT and the MLP (Table~\ref{tab:main} and Fig.~\ref{fig:roc}), with only the four-momentum of each jet as input and $b$ tagging used only to choose the candidate jets.
Without the additional-jet attention, CoReNet uses only the four candidate jets and the event-level quantities and reaches $0.8578 \pm 0.0013$ with 782 parameters, still well above the kinematic baselines but lower than the full network ($p = 0.012$).
In $\bbWW$, CoReNet$_{\HH}$, which encodes only the decay chain of the signal, reaches $0.8859 \pm 0.0016$, against $0.8536 \pm 0.0041$ and $0.8537 \pm 0.0019$ (Table~\ref{tab:bbww-hyp} and Fig.~\ref{fig:bbww-roc}).

\begin{figure}[t]
  \centering
  \includegraphics[width=0.78\textwidth]{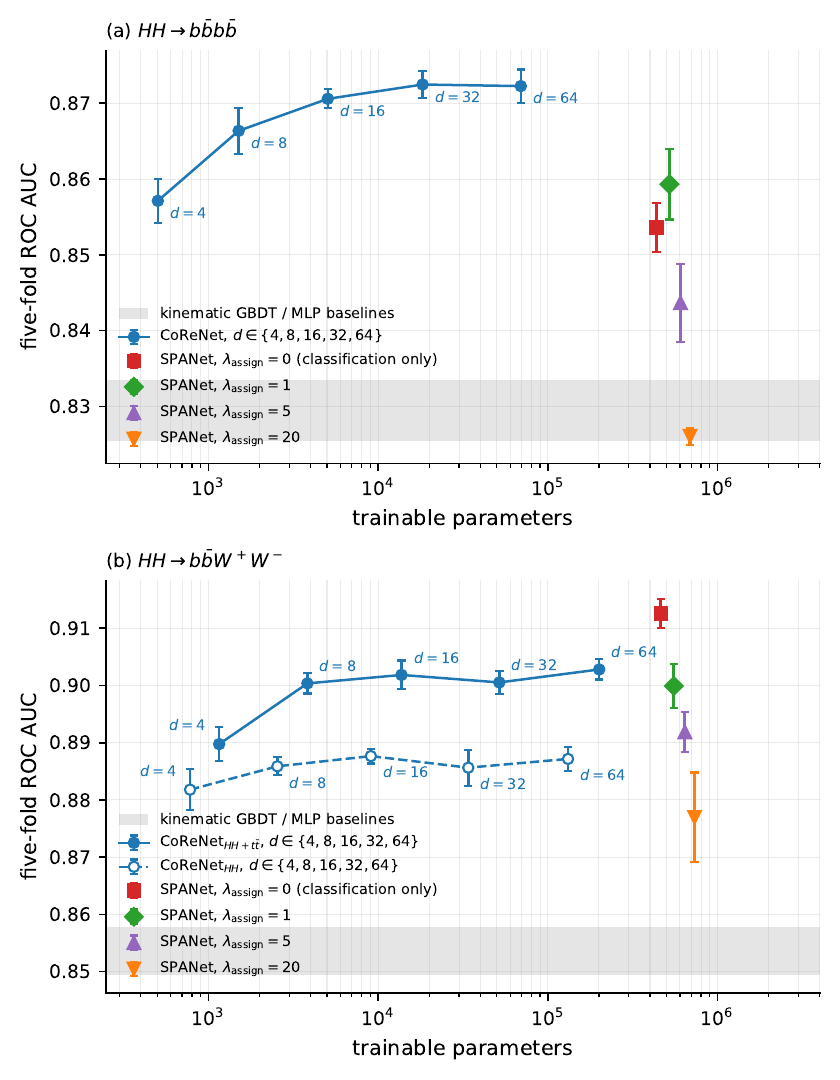}
  \caption{Classification AUC (test) versus trainable parameters for (a) $\HH \to \bbbb$ and (b) $\HH \to \bbWW$. CoReNet is shown at
  $d \in \{4,8,16,32,64\}$ with five-fold standard deviations, in (b) as CoReNet$_{\HH+\ttbar}$ and CoReNet$_{\HH}$; SPANet appears at
  559846 (a) and 594520 (b) parameters at the four assignment-loss weights
  $\lambda_{\mathrm{assign}} = 0$ (classification only, $\lambda_{\mathrm{det}}=0,\ \lambda_{\mathrm{cls}}=1$),
  1, 5 and 20, the latter three with $\lambda_{\mathrm{det}}=\lambda_{\mathrm{cls}}=1$,
  horizontally offset for visibility; the shaded band is the spread of the
  hand-engineered GBDT and MLP baselines.}
  \label{fig:scaling}
\end{figure}

This performance is reached with a network more than two orders of magnitude smaller than the state-of-the-art reference (Fig.~\ref{fig:scaling}).
In $\bbbb$, the best SPANet configuration, trained with $\lambda_{\mathrm{assign}} = 1$, reaches $0.8593 \pm 0.0047$ with 559846 parameters, 371 times more than CoReNet.
CoReNet is ahead in every fold, although with five folds the difference is not significant ($p = 0.14$; all $p$-values use the Nadeau--Bengio correction~\cite{Nadeau:2003}).
Trained for 150 instead of 500 epochs, the same configuration reaches $0.8624 \pm 0.0059$, still below CoReNet (Table~\ref{tab:lambda}).
This holds although SPANet receives the $b$ tag of every jet as an input, which in a $\bbbb$ final state, where the background consists of configurations that mimic four $b$ jets, is a directly discriminating input.
In $\bbWW$, the best SPANet configuration is the one trained with the classification loss alone, at $0.9126 \pm 0.0025$, $0.012$ above CoReNet$_{\HH+\ttbar}$ with 155 times more parameters.
In both final states, large assignment-loss weights lower the AUC of SPANet: with the loss weights of Ref.~\cite{Li:2024lml} ($\lambda_{\mathrm{assign}} = 20$) it reaches $0.8261 \pm 0.0011$ in $\bbbb$, slightly below the kinematic baselines, and $0.8770 \pm 0.0078$ in $\bbWW$.

\label{sec:width_scan}
\begin{table}[t]
  \centering
  \caption{CoReNet width scan on $\HH \to \bbbb$. Only the embedding dimension
  $d$ changes; everything else follows the main training. ROC AUC on the
  training folds, the validation fold and the held-out test fold, five-fold
  mean $\pm$ standard deviation.}
  \label{tab:dscan}
  \small
  \begin{tabular}{crccc}
    \toprule
    $d$ & Parameters & AUC (train) & AUC (validation) & AUC (test) \\
    \midrule
    4  & 504     & $0.8587 \pm 0.0025$ & $0.8575 \pm 0.0039$ & $0.8571 \pm 0.0029$ \\
    \textbf{8}  & \textbf{1508}  & $0.8723 \pm 0.0018$ & $0.8670 \pm 0.0042$ & $\mathbf{0.8664 \pm 0.0031}$ \\
    16 & 5052  & $0.8835 \pm 0.0020$ & $0.8713 \pm 0.0017$ & $0.8706 \pm 0.0013$ \\
    32 & 18284 & $0.8883 \pm 0.0012$ & $0.8726 \pm 0.0018$ & $0.8725 \pm 0.0017$ \\
    64 & 69324 & $0.8887 \pm 0.0046$ & $0.8729 \pm 0.0018$ & $0.8723 \pm 0.0022$ \\
    \bottomrule
  \end{tabular}
\end{table}

\begin{table}[t]
  \centering
  \caption{CoReNet$_{\HH}$ and CoReNet$_{\HH+\ttbar}$ width scans on $\HH \to \bbWW$. Only the embedding dimension
  $d$ changes; everything else follows the main training. ROC AUC on the
  training folds, the validation fold and the held-out test fold, five-fold
  mean $\pm$ standard deviation.}
  \label{tab:dscan-bbww}
  \small
  \begin{tabular}{crccc}
    \toprule
    $d$ & Parameters & AUC (train) & AUC (validation) & AUC (test) \\
    \midrule
    \multicolumn{5}{l}{CoReNet$_{\HH}$} \\
    4  & 780    & $0.8877 \pm 0.0010$ & $0.8812 \pm 0.0027$ & $0.8818 \pm 0.0035$ \\
    \textbf{8}  & \textbf{2550} & $0.8980 \pm 0.0012$ & $0.8863 \pm 0.0020$ & $\mathbf{0.8859 \pm 0.0016}$ \\
    16 & 9066   & $0.9104 \pm 0.0021$ & $0.8878 \pm 0.0015$ & $0.8877 \pm 0.0013$ \\
    32 & 34002  & $0.9139 \pm 0.0102$ & $0.8870 \pm 0.0021$ & $0.8856 \pm 0.0032$ \\
    64 & 131490 & $0.9235 \pm 0.0008$ & $0.8880 \pm 0.0027$ & $0.8872 \pm 0.0021$ \\
    \midrule
    \multicolumn{5}{l}{CoReNet$_{\HH+\ttbar}$} \\
    4  & 1158   & $0.8965 \pm 0.0042$ & $0.8901 \pm 0.0052$ & $0.8898 \pm 0.0030$ \\
    \textbf{8}  & \textbf{3832} & $0.9160 \pm 0.0015$ & $0.9003 \pm 0.0007$ & $\mathbf{0.9004 \pm 0.0018}$ \\
    16 & 13740  & $0.9241 \pm 0.0050$ & $0.9023 \pm 0.0010$ & $0.9018 \pm 0.0025$ \\
    32 & 51796  & $0.9264 \pm 0.0088$ & $0.9016 \pm 0.0018$ & $0.9005 \pm 0.0020$ \\
    64 & 200868 & $0.9288 \pm 0.0063$ & $0.9026 \pm 0.0011$ & $0.9028 \pm 0.0018$ \\
    \bottomrule
  \end{tabular}
\end{table}

A small network is sufficient.
Already at $d = 4$, with 504 parameters, CoReNet reaches $0.8571 \pm 0.0029$ in $\bbbb$, well above the kinematic baselines (Table~\ref{tab:dscan}).
The test AUC is highest at $d = 32$, $0.8725 \pm 0.0017$, and the configuration with $d = 8$ used for the main results has $12.1$ times fewer parameters and a test AUC lower by 0.7\%.
Above $d = 8$ the test AUC keeps rising and plateaus at $d = 32$, with $10^{4}$ parameters.
The $\bbWW$ scans gain less from width (Table~\ref{tab:dscan-bbww}, Fig.~\ref{fig:scaling}b): for both configurations the test AUC changes by less than 0.3\% from $d = 8$ to $d = 64$.
At $d = 4$ the test AUC is lower than at $d = 8$ by 0.5\% (CoReNet$_{\HH}$) and 1.2\% (CoReNet$_{\HH+\ttbar}$).
The gap between the training and validation AUC increases with width, from $0.006$ at $d = 4$ to $0.026$ at $d = 64$ for CoReNet$_{\HH+\ttbar}$, and from $0.007$ to $0.036$ for CoReNet$_{\HH}$.
The $\ttbar$ hypothesis adds $0.014$--$0.016$ to the test AUC from $d = 8$ upward.
In $\bbWW$ the $d = 8$ configuration, with $2.5 \times 10^{3}$ and $3.8 \times 10^{3}$ parameters for CoReNet$_{\HH}$ and CoReNet$_{\HH+\ttbar}$, is within 0.3\% of every larger width, and every width is well above the kinematic baselines; SPANet trained with the classification loss alone is $0.012$ higher with 155 times more parameters, and the assignment loss lowers its AUC (Fig.~\ref{fig:scaling}b).

Encoding the decay chain of the background adds information that the signal hypothesis alone does not provide.
In $\bbWW$, CoReNet$_{\HH+\ttbar}$ reaches $0.9004 \pm 0.0018$, higher than CoReNet$_{\HH}$ in every fold ($p = 3 \times 10^{-5}$), at the cost of 1282 additional parameters.
On its own, the background hypothesis is a weaker discriminant: the auxiliary $\ttbar$ output reaches $0.8753 \pm 0.0005$, well below the main output, so the two hypotheses are most useful together.

The hypothesis weights follow the decay structure of the signal, although CoReNet is not trained to find the truth-matched pairing.
The weights $\qscore$ are whatever weighting of the hypotheses best serves the classification (Sec.~\ref{sec:architecture}), and a weighting that differs from the truth-matched pairing is not an error, but it is instructive to compare them with it, on the $\bbbb$ signal events in which all four $b$ quarks are matched to the four candidate jets.
The hypothesis with the largest $\qscore$, taken as the pairing chosen by CoReNet, agrees with the truth-matched pairing in $80.8 \pm 11.9\%$ of these events, compared with $33.3\%$ for a random choice and $85.2 \pm 0.5\%$ for SPANet trained with $\lambda_{\mathrm{assign}} = 20$, whose assignment loss targets exactly this pairing.
The agreement varies strongly between folds, from 66 to $93\%$, since nothing in the training constrains it, while the classification performance is stable across folds.

Finally, CoReNet trains in minutes.
On one NVIDIA V100 GPU, a fold trains in about 6 minutes for $\bbbb$, including data loading, and in 1 to 5 minutes for $\bbWW$, whereas the 500-epoch SPANet trainings take 2.8--3.9 and 2.0--2.8 hours, excluding the scan of validation checkpoints that runs in the same job.
The difference comes mainly from the number of epochs: CoReNet is trained for 20 and SPANet for the 500 of Ref.~\cite{Li:2024lml}, which the configurations with $\lambda_{\mathrm{assign}} = 5$ and 20 need, since in $\bbbb$ they reach their best validation AUC only after about 407 and 465 epochs on average, and $\lambda_{\mathrm{assign}} = 1$ after about 100 in four of the five folds (App.~\ref{app:spanet}).
The classification-only SPANet reaches its best validation AUC after about 29 epochs; trained for 50 epochs, with a comparable AUC (Table~\ref{tab:lambda}), it takes 16 to 23 minutes per fold for $\bbbb$, excluding the checkpoint scan.

\section{Discussion and conclusions}
\label{sec:discussion}
\label{sec:conclusion}

We have presented CoReNet, a network that builds the known decay chain of a final state into its architecture.
Its event representation follows the hierarchy of the decay, every assignment hypothesis is evaluated with weights shared across objects and combined with learned weights, and the exchange symmetries of the decay are built into the construction.
The network is trained on the classification objective alone and uses no generator-level information.

CoReNet was first described in a study of the data-driven modeling of the multijet background to $\HH \to \bbbb$~\cite{Manole:2022bhs}, and has since been used in the CMS search for $ZZ$ and $ZH$ production in the $\bbbb$ final state~\cite{CMS:2024zzb} and in the CMS search for non-resonant $\HH \to \bbbb$ production~\cite{CMS:HIG24010}.
In the two CMS searches it served both as the final signal-versus-background discriminant and as a density-ratio estimator for the background model, under conditions considerably more demanding than a fast-simulation benchmark, with data-driven backgrounds and systematic uncertainties.
A dedicated study of its performance as a density-ratio estimator will be presented in a separate publication.
That the same architecture has been used successfully in both roles indicates that its representation is not tuned to a single task.

On $\HH \to \bbbb$ against QCD multijet production, CoReNet with 1508 trainable parameters reaches a ROC AUC of $0.8664 \pm 0.0031$, well above kinematic baselines, including an MLP with 4993 parameters, and comparable to or better than every SPANet configuration we tested, the best of which at the 500-epoch budget of the main comparison reaches $0.8593 \pm 0.0047$ with 371 times more parameters and per-jet $b$-tagging information that CoReNet does not receive.
Its test AUC saturates at a width of $d \approx 32$--64 (Table~\ref{tab:dscan}), less than 1\% above the $d = 8$ configuration used throughout, and it trains in about 6 minutes per fold on a single GPU.
Its hypothesis weights select the truth-matched pairing in $81\%$ of the signal events in which it is defined, although no assignment target enters the training.

The construction is not specific to $\HH \to \bbbb$.
For $\HH \to \bbWW$ against semileptonic $\ttbar$, a CoReNet that encodes the decay chain of the signal reaches an AUC of $0.8859 \pm 0.0016$ with 2550 parameters, well above the kinematic baselines.
Encoding the decay chain of the background as well raises it to $0.9004 \pm 0.0018$ with 3832 parameters, $0.012$ below the best SPANet configuration, which has 155 times more parameters.
Hypotheses for the background arise naturally when the decay chains are part of the architecture, and they improve the classification at a cost of about $10^{3}$ parameters.

In both final states, training SPANet with a strong jet-assignment loss lowers its classification performance; in $\bbbb$ it falls slightly below the kinematic baselines, although it then reproduces the truth-matched pairing more often on average than CoReNet does (85.2\% versus 80.8\%).
A jet assignment is useful to an analysis only to the extent that it improves the separation of signal from background, and a network that encodes the decay structure in its architecture needs no such balancing.

Where reconstructing the jet assignment is itself the goal, symmetry-preserving attention networks such as SPANet remain well suited~\cite{Fenton:2020woz,Fenton:2023jdb}.
For classification, a network whose architecture encodes the decay chain reaches comparable performance with more than two orders of magnitude fewer parameters, trains in minutes, and needs no truth-level information.

These results bear on a broader question in machine learning: whether built-in domain knowledge is preferable to larger, more general models trained on more data~\cite{Sutton:2019,Kaplan:2020,Hoffmann:2022}.
At the sample sizes studied here, which are typical of simulated signal samples and of the control regions used in data-driven background estimates, encoding the decay chain in the architecture reaches a performance comparable to that of a network more than two orders of magnitude larger.
In $\HH \to \bbbb$, larger CoReNet widths raise the test AUC relative to $d = 8$ by at most 0.7\% (Table~\ref{tab:dscan}).
In $\bbWW$, they change it by less than 0.3\% (Table~\ref{tab:dscan-bbww}).

Whether the advantage in network size and training time persists with much larger training samples is not tested here; studying the performance of both approaches as a function of the sample size would answer this directly.

The small size of the network also changes where it can be used.
With about $10^{3}$ parameters, the network is small enough to be considered for environments in which a transformer with $10^{5}$--$10^{6}$ parameters is impractical, such as trigger-level selection and inference on field-programmable gate arrays~\cite{Duarte:2018ite}, although its latency there remains to be studied.
Efficiency is therefore not only a matter of computing cost; it determines where a model can run~\cite{Hooker:2021}.

\section*{Acknowledgements}
This work used Bridges-2 at Pittsburgh Supercomputing Center through allocation PHY260026 from the Advanced Cyberinfrastructure Coordination Ecosystem: Services \& Support (ACCESS) program, which is supported by National Science Foundation grants \#2138259, \#2138286, \#2138307, \#2137603, and \#2138296.
The preparation of this manuscript and of the accompanying code repository was assisted by large language models.
All results, text and code were reviewed and verified by the authors, who take full responsibility for the content.

\paragraph{Code and data availability}
The CoReNet implementation, training and evaluation code, and the benchmark dataset preparation chain are publicly available at \url{https://github.com/cms-cmu/CoReNet}~\cite{CoReNet:software}.
The simulated samples are available at Ref.~\cite{Khanal:2026data}.

\paragraph{Funding information}
This work was supported by the U.S. Department of Energy, Office of Science, Office of High Energy Physics, under Award No.~DE-SC0010118.
A.G.E. is supported in part by the Fermilab LHC Physics Center (LPC) Distinguished Researcher program.

\begin{appendix}
\numberwithin{equation}{section}
\section{CoReNet implementation details}
\label{app:corenet}

This appendix gives the details of the CoReNet implementation that are needed to reproduce the network but not to follow its design, which is described in Sec.~\ref{sec:architecture}.
The public implementation, \url{https://github.com/cms-cmu/CoReNet}, is based on PyTorch~\cite{Paszke:2019xhz} and follows this description.
Unless stated otherwise, the details are given for $\HH \to \bbbb$; the differences for $\HH \to \bbWW$ are listed at the end of this appendix.

The transverse momenta and masses of all objects enter as $\log(1+\pt)$ and $\log(1+m)$, and the jet multiplicity as $\log(N_{\mathrm{jets}} - (N_{\mathrm{min}} - 1))$, where $N_{\mathrm{min}}$ is the minimum number of selected jets, so that the minimum multiplicity maps to zero; both final states require at least four jets, $N_{\mathrm{min}} = 4$.
Every embedding layer is preceded by ghost batch normalization~\cite{Hoffer:2017}, whose statistics are initialized on the training sample.
These input transformations are the same in both final states.
In the CMS analyses, the event-level quantities also include the data-taking year and quantities related to top quark decays.

The six dijets are ordered as $(12), (34), (13), (24), (14), (23)$, so that consecutive pairs form the three pairing hypotheses.
Each jet enters once for every dijet it belongs to, with its azimuth measured relative to that dijet,
\begin{equation}
  \mathbf{x}_{i}^{(ij)} = \big(\log(1+\pt^{\,i}),\; \eta_i,\;
    |\Delta\phi_{i,ij}|,\; \log(1+m_i)\big),
\end{equation}
and each dijet is described by the kinematics of its four-vector sum, with its azimuth measured relative to the four-jet system,
\begin{equation}
  \mathbf{y}_{ij} = \big(\log(1+\pt^{\,ij}),\; \eta_{ij},\;
    |\Delta\phi_{ij,4j}|,\; \log(1+m_{ij})\big).
\end{equation}

\begin{figure}[tbp]
  \centering
  \includegraphics[width=\textwidth]{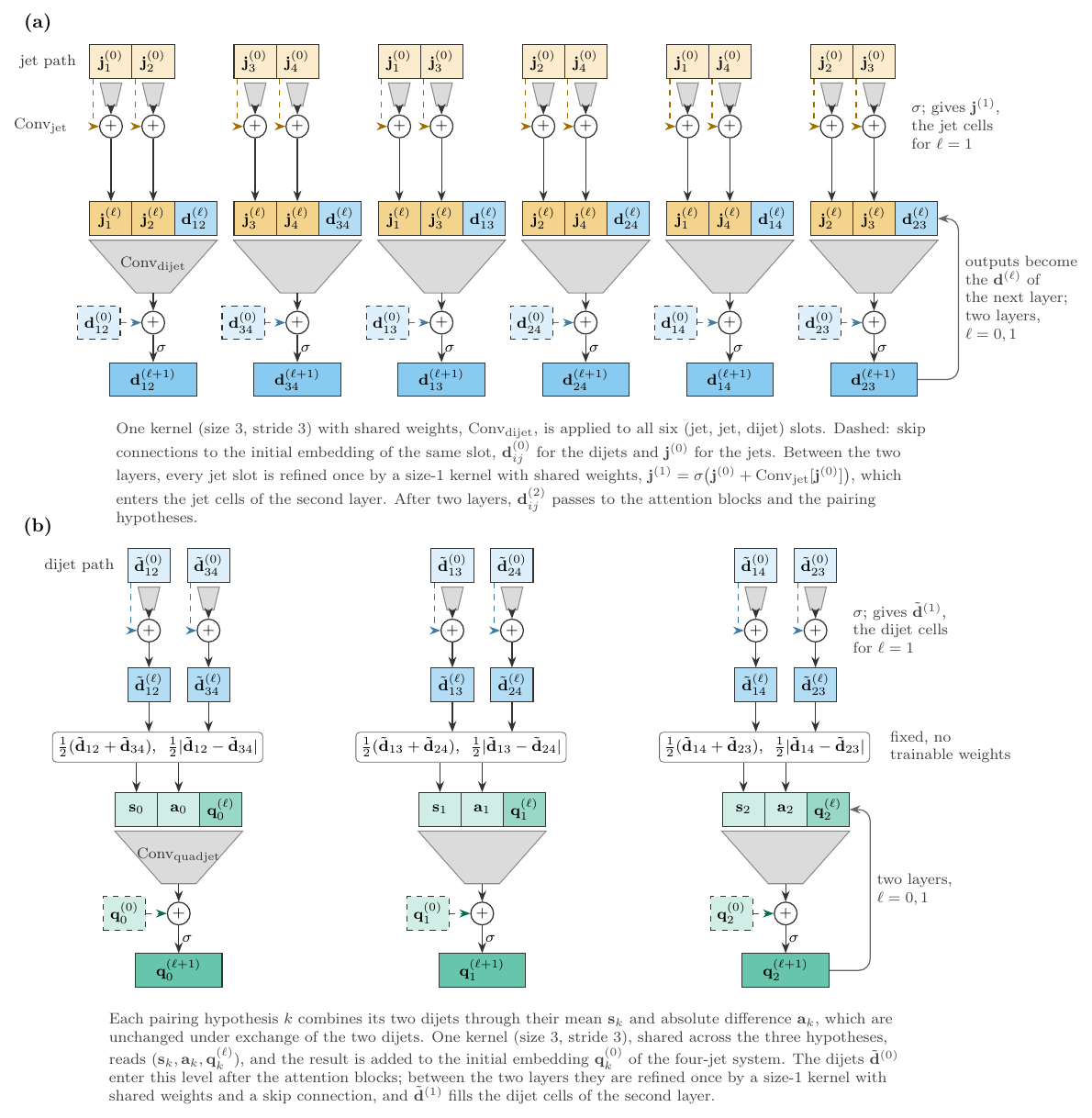}
  \caption{The residual blocks of (a) the dijet level, Eq.~\eqref{eq:dijet-block}, and (b) the pairing-hypothesis level, Eq.~\eqref{eq:quadjet-block}.
  (a) The jets and dijets are laid out as six slots of (jet, jet, dijet), and a single convolution with kernel size and stride three, with the same weights for every slot, maps each slot to a dijet update.
  Each update is added to the initial embedding $\mathbf{d}^{(0)}_{ij}$ of the same slot and passed through the SiLU nonlinearity $\sigma$.
  The outputs replace $\mathbf{d}^{(\ell)}_{ij}$ in the next layer, and two layers are applied.
  Between them, each jet slot is refined once by a size-1 convolution with a skip connection to $\mathbf{j}^{(0)}$ (jet path), and the second layer reads the refined jets $\mathbf{j}^{(1)}$.
  (b) For each pairing hypothesis, the two dijets are combined through their mean $\mathbf{s}_k$ and absolute difference $\mathbf{a}_k$, which carry no trainable weights and are unchanged under exchange of the two dijets.
  A single convolution, shared across the three hypotheses, reads $(\mathbf{s}_k, \mathbf{a}_k, \mathbf{q}_k^{(\ell)})$, and its output is added to the initial embedding $\mathbf{q}_k^{(0)}$ of the four-jet system.
  The dijets $\tilde{\mathbf{d}}$ entering this level are refined once between the two layers in the same way as the jets in (a) (dijet path).}
  \label{fig:dijet-block}
\end{figure}

The residual blocks of the dijet and hypothesis levels, Eqs.~\eqref{eq:dijet-block} and~\eqref{eq:quadjet-block}, are shown in Fig.~\ref{fig:dijet-block}.

\begin{figure}[t]
  \centering
  \includegraphics[width=\textwidth]{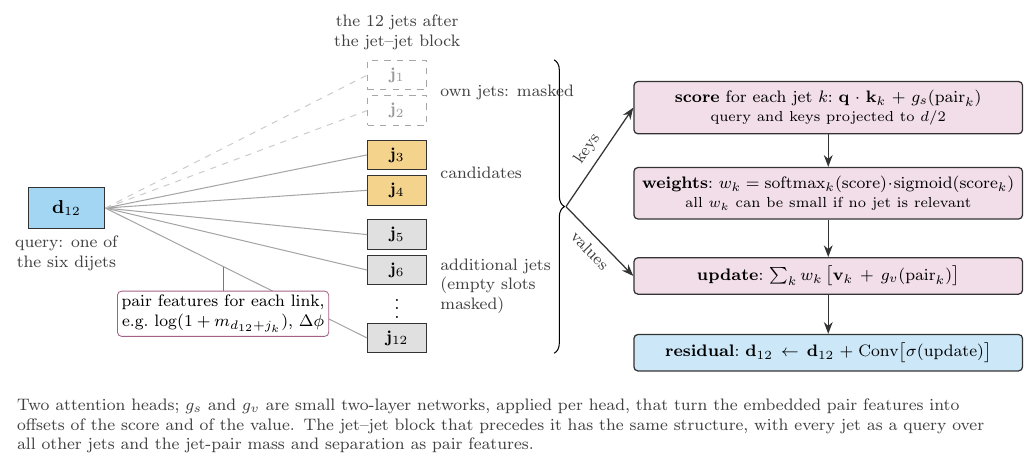}
  \caption{The dijet--jet attention block, shown for the dijet $\mathbf{d}_{12}$ as the query.
  The values are the twelve jets after the jet--jet block, with the two jets of the dijet and the empty slots masked.
  For each query--jet pair, the logarithm of the invariant mass and the azimuthal separation are embedded and enter the score and the value through small networks $g_s$ and $g_v$.
  The attention weights are the softmax of the scores multiplied by their sigmoid, and the weighted sum of the values updates the dijet as a residual correction.
  The jet--jet block that precedes it has the same structure.}
  \label{fig:attention}
\end{figure}

The additional jets enter through two attention blocks with two heads each (Fig.~\ref{fig:attention}).
All twelve jets (the four candidates and up to eight others) are embedded from $(\log(1+\pt),\allowbreak \eta,\allowbreak \log(1+m))$ and their candidate label, again with the event-level embedding added.
For every pair of objects, the logarithm of the invariant mass and the azimuthal separation are embedded and used to modify both the attention score and the value.

In the first block, each jet attends to all other jets, with the jet-pair mass and separation as pair features.
In the second, each of the six dijets attends to the twelve jets, excluding its own two, with the mass of the dijet--jet system and their azimuthal separation as pair features.
In both blocks, the pair features pass through a two-layer perceptron per head, whose output is added to the score and to the value.
Queries and keys are projected to a reduced dimension of $d/2$ before the score is computed, while the values keep the full width $d$.
The scores are normalized with ghost batch normalization, and the attention weights are the softmax of the scores multiplied by their sigmoid, so a query can assign low weight to every value when none is relevant.
Empty jet slots are masked.
The output of the second block is added to the dijet representation, again as a residual correction, before the hypothesis level.

In the pooling of Eq.~\eqref{eq:pooling}, the scoring layer has no bias: a bias would have no effect, since the softmax is unchanged when a constant is added to all scores.

CoReNet is trained with the recipe of its CMS deployments~\cite{CMS:2024zzb,CMS:HIG24010}.
The optimizer is Adam~\cite{Kingma:2014} with a learning rate of $10^{-2}$, and training runs for 20 epochs, after which the epoch with the best validation AUC is kept.
The batch size starts at 1024 and is doubled after epochs 1, 3, 6 and 10, while the number of ghost batches is reduced accordingly.
After epoch 15, all layers except the output layer are frozen and the learning rate is reduced by a factor of four at every epoch.

For the width scan of Sec.~\ref{sec:width_scan}, $d = 4$, 16, 32 and 64 give 504, 5052, 18284 and 69324 parameters.

For $\HH \to \bbWW$, the merge of the two $W$ candidates, and of the two top quark candidates in CoReNet$_{\HH+\ttbar}$, is a convolution that reads the two candidates together with the invariant mass and the angular separations of the pair.
The three relation features are the invariant mass and the angular separations of the lepton and the missing transverse momentum, of the $H \to b\bar{b}$ candidate and the missing transverse momentum, and of the lepton and the hadronic $W$ candidate.

\section{SPANet training details}
\label{app:spanet}

This appendix documents the SPANet reference of Sec.~\ref{sec:setup}: how it is trained, how its assignment targets are defined, and how its classification performance depends on the weight of the assignment loss and on the number of training epochs.
We use the architecture and optimizer settings of Ref.~\cite{Li:2024lml} and, as there, train for 500 epochs, on training samples roughly ten times smaller than theirs.
For each fold, the checkpoint with the best validation AUC is used, as for CoReNet.
Table~\ref{tab:training} compares the two training configurations.
Both networks are trained and evaluated on the same events, folds and class-normalized event weights, and they differ in how they receive the $b$ tag: CoReNet only through the choice and ordering of the candidate jets, SPANet as a per-jet input feature.
For $\HH \to \bbWW$, SPANet uses the same architecture and training settings.
In addition to the per-jet features, it receives the lepton ($\pt$, $\eta$, $\phi$, $m$ and its flavor) and the missing transverse momentum ($\pt$, $\phi$) as event-level inputs, and it has 594520 trainable parameters.
Its assignment targets are the two jets from the $H \to b\bar{b}$ decay and the two jets from the hadronic $W$ decay, each symmetric under exchange of its two jets, and its classification target is signal versus $\ttbar$.

\begin{table}[t]
  \centering
  \caption{Training configurations for $\HH \to \bbbb$. Both networks use the same folds, the same
  class-normalized event weights and the same input file; the $b$ tag enters CoReNet
  only through the choice and ordering of the candidate jets, and SPANet as a per-jet
  feature. SPANet's architecture and optimizer follow Ref.~\cite{Li:2024lml}.}
  \label{tab:training}
  \footnotesize
  \begin{tabular}{@{}l p{4.6cm} p{5.9cm}@{}}
    \toprule
    & CoReNet ($d = 8$) & SPANet \\
    \midrule
    Folds                  & \multicolumn{2}{l}{5 (3 training, 1 validation, 1 test)} \\
    Trainable parameters   & 1508 & 559846 \\
    Width                  & $d = 8$ & hidden dimension 32, 8 attention heads \\
    Depth                  & jet $\to$ dijet $\to$ quadjet hierarchy & 8 central + 6 branch encoder layers \\
    Per-jet inputs         & $\pt,\ \eta,\ \phi,\ m$ & $\pt,\ \eta,\ \phi,\ m$, $b$ tag \\
    Optimizer              & Adam, learning rate $10^{-2}$ & AdamW~\cite{Loshchilov:2019}, learning rate $1.5 \times 10^{-3}$ \\
    Learning-rate schedule & $\times 0.25$ per epoch after epoch 15 & one cosine-annealing cycle~\cite{Loshchilov:2016tgk} \\
    Batch size             & 1024, doubled after epochs 1, 3, 6, 10 & 4096 \\
    Regularization         & none & dropout 0.2, gradient clipping 1.0, $L_2$ penalty $2 \times 10^{-4}$ \\
    Epochs                 & 20 & 500 \\
    Late-training rule     & only output layer trained after epoch 15 & --- \\
    Model selection        & epoch with the best validation AUC & epoch with the best validation AUC \\
    Loss                   & classification cross-entropy & $\lambda_{\mathrm{assign}}\mathcal{L}_{\mathrm{assign}} + \lambda_{\mathrm{det}}\mathcal{L}_{\mathrm{det}} + \lambda_{\mathrm{cls}}\mathcal{L}_{\mathrm{cls}}$ \\
    \bottomrule
  \end{tabular}
\end{table}

The assignment and detection losses of SPANet require generator-level targets, the jets that originate from each parent particle.
For $\HH \to \bbbb$, they are obtained by matching the $b$ quarks from the Higgs boson decays in signal events to the selected jets within $\Delta R < 0.4$, using a globally optimal assignment~\cite{Kuhn:1955}.
All four quarks are matched to distinct jets in $45.0\%$ of the selected signal events.
For $\HH \to \bbWW$, the two $b$ quarks from the Higgs boson decay and the two quarks from the hadronic $W$ decay are matched to distinct jets in the same way; all four are matched in $21.1\%$ of the selected signal events (3697 events).
These matches are used to train SPANet and, for $\bbbb$, to compare the hypothesis weights of CoReNet with the truth-matched pairing (Sec.~\ref{sec:results}); CoReNet is never trained on them.

To check that the comparison does not depend on the training budget, each of the four loss configurations of Sec.~\ref{sec:setup} is also trained on $\HH \to \bbbb$ for 50 and 150 epochs, and the results for the three budgets are given in Table~\ref{tab:lambda}.

\begin{table}[t]
  \centering
  \caption{SPANet classification AUC on $\HH \to \bbbb$ as a function of the assignment-loss weight
  $\lambda_{\mathrm{assign}}$ ($\lambda_{\mathrm{det}} = \lambda_{\mathrm{cls}} = 1$, except
  $\lambda_{\mathrm{det}} = 0$ for the classification-only row), for training budgets of 50, 150 and 500 epochs.
  Each entry is the five-fold mean $\pm$ standard deviation of the test AUC at the best-validation checkpoint; the
  last column gives the mean best-validation epoch of the 500-epoch runs.
  For comparison, CoReNet ($d=8$) reaches $0.8664 \pm 0.0031$.}
  \label{tab:lambda}
  \footnotesize
  \begin{tabular}{lrrrr}
    \toprule
    $\lambda_{\mathrm{assign}}$ & 50 epochs & 150 epochs & 500 epochs & best epoch \\
    \midrule
    0 (classification only) & $0.8555 \pm 0.0017$ & $0.8545 \pm 0.0035$ & $0.8536 \pm 0.0032$ & 29 \\
    1                       & $0.8520 \pm 0.0058$ & $0.8624 \pm 0.0059$ & $0.8593 \pm 0.0047$ & 150 \\
    5                       & $0.8338 \pm 0.0064$ & $0.8361 \pm 0.0051$ & $0.8437 \pm 0.0052$ & 407 \\
    20                      & $0.8075 \pm 0.0075$ & $0.8188 \pm 0.0106$ & $0.8261 \pm 0.0011$ & 465 \\
    \bottomrule
  \end{tabular}
\end{table}

The classification-only SPANet reaches its highest validation AUC after about 29 epochs and $\lambda_{\mathrm{assign}} = 1$ after about 100 in four of the five folds, whereas $\lambda_{\mathrm{assign}} = 5$ and 20 keep improving until late in the 500-epoch training (about 407 and 465 epochs), so shorter budgets understate them.
Large assignment-loss weights lower the AUC: with the loss weights of Ref.~\cite{Li:2024lml} ($\lambda_{\mathrm{assign}} = 20$, $\lambda_{\mathrm{det}} = 1$, $\lambda_{\mathrm{cls}} = 1$) it falls to $0.8261 \pm 0.0011$, slightly below the kinematic baselines, and $\lambda_{\mathrm{assign}} = 5$ gives $0.8437 \pm 0.0052$.
In $\bbbb$, a small weight, $\lambda_{\mathrm{assign}} = 1$, is slightly better than none ($+0.006$, $p = 0.023$), so a weak hint of the decay structure helps the transformer there, whereas in $\bbWW$ the classification-only configuration is the best (Table~\ref{tab:bbww-hyp}); in both final states the configuration trained hardest to reproduce the truth-matched pairing has the lowest AUC.
Reproducing the truth-matched pairing and separating signal from background are therefore related but different tasks, and CoReNet, which receives the decay structure through its architecture rather than through its loss, has no such trade-off to tune.

\end{appendix}

\bibliography{references}

\end{document}